\documentclass[
 jcp,
 amsmath,
 amssymb,
superscriptaddress,
preprint,
]{revtex4-1}

\usepackage[colorlinks=true, linkcolor=green, citecolor=red]{hyperref}

\usepackage{graphicx}
\usepackage{dcolumn}
\usepackage{bm}
\usepackage[utf8]{inputenc}
\usepackage{textcomp}
\usepackage{amsmath}
\usepackage{color}
\usepackage[version=4]{mhchem}
\usepackage{chemstyle}
\usepackage{siunitx}
\usepackage{textgreek}
\usepackage{booktabs}

\usepackage{booktabs}
\usepackage{multirow}
\usepackage{threeparttable} 

\usepackage{gensymb}

\begin{document}

\title{
On the role of nuclear quantum effects on the stability of peptides
}

\author{Jing Shen}
\affiliation{Department of Chemistry, Shanghai Key Laboratory of Molecular Catalysis and Innovative Materials, State Key Laboratory of Porous Materials for Separation and Conversion, Fudan University, Shanghai 200438, P. R. China}
\author{Ming-Zheng Du}
\affiliation{Department of Chemistry, Shanghai Key Laboratory of Molecular Catalysis and Innovative Materials, State Key Laboratory of Porous Materials for Separation and Conversion, Fudan University, Shanghai 200438, P. R. China}
\author{Dong H. Zhang}
\affiliation{State Key Laboratory of Chemical Reaction Dynamics, Dalian Institute of Chemical Physics, Chinese Academy of Sciences, Dalian 116023, P. R. China}
\affiliation{ University of Chinese Academy of Sciences, Beijing 100049, China }
\author{Venkat Kapil}
\email{v.kapil@ucl.ac.uk}
\affiliation{Department of Physics and Astronomy, London Centre for Nanotechnology, Thomas Young Centre, University College London, WC1E 6BT London, United Kingdom}
\author{Wei Fang}
\email{wei\_fang@fudan.edu.cn}
\affiliation{Department of Chemistry, Shanghai Key Laboratory of Molecular Catalysis and Innovative Materials, State Key Laboratory of Porous Materials for Separation and Conversion, Fudan University, Shanghai 200438, P. R. China}

\begin{abstract}

Nuclear quantum effects (NQEs) arising from the light mass of hydrogen can influence the structure and stability of hydrogen-bonded biomolecules, yet their role in determining peptide and protein folding remains unclear. 
Experiments show that substituting \ce{H2O} with \ce{D2O} often stabilizes folded states, but the microscopic mechanism associated with this phenomena remains unresolved. 
Through \textit{ab initio}–level path-integral molecular dynamics simulations enabled by machine-learning interatomic potentials, we address the fundamental question of the role of NQEs in peptides by investigating both their overall impact and isotope substitution effects.
Overall,  NQEs systematically destabilize compact three-dimensional structures across peptide systems, independent of secondary structure type or side-chain interactions. 
Contrary to the conventional picture that places central importance on hydrogen bonds, we find that the dominant destabilization instead arises from the quantum C--H vibrations. 
In addition, we reveal microscopic insights into the stabilization of folded peptides upon \ce{H2O} to \ce{D2O} substitution, showing that the H/D isotope substitution of active peptide hydrogens, previously considered unimportant, produces free-energy changes within the range of experimentally observed shifts. 
These findings provide a new interpretation of isotope effects in biological systems, indicating that seemingly small H $\to$ D substitutions within peptides can be as important as, or even outweigh, solvent contributions.

\end{abstract}

\maketitle

\section{Introduction}
The folding of peptide sequences into three-dimensional structures is fundamental to protein function~\cite{Anfinsen_SciNewYorkNY_1973_v181_p223,Dobson_Nature_2003_v426_p884,Jumper_Nature_2021_v596_p583}.
This transformation from disordered coils (unfolded) to ordered structures (folded) is governed by a delicate balance of non-covalent interactions, such as hydrogen bonding (HB), van der Waals forces, electrostatics, and hydrophobic effects~\cite{Dill_Biochemistry_1990_v29_p7133, Newberry_AcsChemBiol_2019_v14_p1677}.
Among these, HBs deserve particular attention due to their directional nature and crucial role in stabilizing secondary structures~\cite{PAULING_ProcNatlAcadSciUSA_1951_v37_p205,Pauling_ProcNatlAcadSciUSA_1951_v37_p729} in gaseous~\cite{Hudgins_JAmChemSoc_1999_v121_p3494} and solvent phases~\cite{Rose_AnnuRevBiophysBiomolStruct_1993_v22_p381,Pace_ProteinSciPublProteinSoc_2014_v23_p652}.

Due to the light mass of hydrogen atoms that form them, 
the structure and properties of HBs are strongly modulated by quantum nuclear motion.~\cite{Li_ProcNatlAcadSciUSA_2011_v108_p6369}
Nuclear quantum effects (NQEs) in biomolecules have profound real-world implications.
A prominent example is the burgeoning field of deuterated drugs—an FDA-approved strategy attracting multibillion-dollar commercial interest—where deuteration influences not only the metabolism of drug molecules but also all ADMET parameters (absorption, distribution, metabolism, excretion, and toxicity)~\cite{DiMartino_NatRevDrugDiscov_2023_v22_p562}.
Multiple deuterium substitutions can cumulatively reduce plasma protein binding~\cite{Cherrah_BiomedEnvironMassSpectrom_1987_v14_p653}, alter absorption~\cite{Parente_FoodChemToxicolIntJPublBrIndBiolResAssoc_2022_v160_p112774}, and affect lipophilicity~\cite{el1984lipophilicity,Bechalany_HelvChimActa_1989_v72_p472}.
At the organismal level, high \ce{D2O} concentrations impair growth and survival across diverse taxa, from microorganisms and algae to plants and mammals, while in mammalian cells, \ce{H2O} to \ce{D2O} substitution at $\sim$45\% markedly suppresses epithelial migration and proliferation -- an effect with emerging applications in organ preservation and anticancer strategies~\cite{Schnauss_AdvMater_2021_v33_p2170230,Jandova_Cancers_2021_v13_p605}.

Isotope effects in the thermodynamics of biomolecules serve as a probe into the origins of these phenomena.
Experiments observe that changing the solvent isotopomers from \ce{H2O} to \ce{D2O} alters the folding equilibrium of peptides~\cite{Giubertoni2023}, predominately stabilizing the folded state~\cite{Parker_Biochemistry_1997_v36_p5786,Chellgren_JAmChemSoc_2004_v126_p14734,Cho_JAmChemSoc_2009_v131_p15188}, with rare cases of unfolded-state stabilization~\cite{Makhatadze_NatStructBiol_1995_v2_p852}.
These observations have been attributed to the enhanced hydrophobic protein–water interactions in \ce{D2O}~\cite{Giubertoni2023}, as
\ce{D2O} possesses stronger HBs than \ce{H2O}~\cite{Ceriotti_ChemRev_2016_v116_p7529}.
In contrast, an alternate explanation—based on the impact of NQEs on the intrinsic stability of proteins arising from H$\to$D substitution of active peptide hydrogen atoms (which exchange via the solvent)—has been interpreted to be negligible~\cite{Parker_Biochemistry_1997_v36_p5786}.
This view has been recently reinforced by a recent gas-phase experiment showing virtually identical backbone structures for deuterated and protonated proteins~\cite{Haidar_JAmSocMassSpectrom_2023_v34_p1447}.
However, unambiguously disentangling the different contributions to free energy differences is very difficult experimentally.
Consequently, the microscopic origins of isotope effects on peptide stability remain elusive.
The lack of a clear fundamental understanding hinders the rational prediction and engineering of isotopic substitutions in applications.

In this context, first-principles simulations that explicitly incorporate NQEs therefore offer a powerful tool for understanding the microscopic origins of isotope effects on peptide stability, and furthermore, shed light on the fundamental question on the role of NQEs in biomolecules.
NQEs have attracted broad interest, with theoretical investigations across diverse systems employing a combination of path-integral simulations and first-principles calculations,
including water~\cite{Benoit_Nature_1998_v392_p258, Chen_PhysRevLett_2003_v91_p215503, Morrone_PhysRevLett_2008_v101_p17801, Paesani_JPhysChemB_2009_v113_p5702}, biomolecules~\cite{Wang_ProcNatlAcadSciUSA_2014_v111_p18454, Rossi2015,Fang_JPhysChemLett_2016_v7_p2125}, molecular crystals~\cite{Rossi_PhysRevLett_2016_v117_p115702, Kapil_ProcNatlAcadSciUSA_2022_v119_pe2111769119, DellaPia2025}, organic liquids~\cite{Pereyaslavets_ProcNatlAcadSciUSA_2018_v115_p8878,Ugur2025}, and superconducting systems \cite{Errea_Nature_2016_v532_p81}.
NQEs in HBs mainly fall into two categories: primary and secondary effects.
The former refers to enhanced fluctuations arising from proton delocalization within the H-bond. 
Prominent examples include enhanced water autoionization~\cite{Ceriotti_ChemRev_2016_v116_p7529}, proton sharing in interfacial water~\cite{Li_PhysRevLett_2010_v104_p66102} and high-pressure ice~\cite{Benoit_Nature_1998_v392_p258, Goncharov_SciNewYorkNY_1996_v273_p218}, and, in the context of biological systems, significant shifts in enzyme pK\textsubscript{a}~\cite{Wang_ProcNatlAcadSciUSA_2014_v111_p18454}.
Secondary effects induce global structural changes and/or thermodynamic shifts, such as the well-known Ubbelohde effect~\cite{Ubbelohde_ActaCryst_1955_v8_p71,Shi_NatCommun_2018_v9_p481}.
These phenomena have been investigated and rationalized in liquid water, small gas-phase clusters, simple molecular crystals
\cite{Habershon_JChemPhys_2009_v131_p24501,Li_ProcNatlAcadSciUSA_2011_v108_p6369,Markland_ProcNatlAcadSciUSA_2012_v109_p7988,Romanelli_JPhysChemLett_2013_v4_p3251,Ceriotti_ProcNatlAcadSciUSA_2013_v110_p15591,Rossi2015,Fang_JPhysChemLett_2016_v7_p2125,Cheng_JPhysChemLett_2016_v7_p2210,Rossi_PhysRevLett_2016_v117_p115702}.
However, due to the high cost of first-principles path-integral molecular dynamics (PIMD) simulations that can account for these effects, secondary NQEs beyond simple systems--such as their impact on the thermodynamic stability of complex, more realistic biomolecules--remain largely unexplored and challenging to elucidate.

In this work, we quantify the impact of NQEs on peptides from first-principles using a quantum accurate state-of-the-art framework.
We combines machine learning interatomic potentials (MLIPs) for efficient and accurate quantum-mechanical-level interactions, PIMD for incorporating NQEs, and thermodynamic integration for rigorous evaluations of free energy differences. 
We first estimate the change in Helmholtz free energy (thermodynamic stability under constant $T$ and $V$ conditions) between the folded and unfolded states of proteins. 
For gas-phase peptides, this is equal to the Gibbs free energy difference relevant for experimental conditions.
To strengthen the connection with experiments, we then quantify the free-energy change upon \ce{H} $\to$ \ce{D} substitution of the active hydrogen in gas-phase peptides. 
Our conclusions are twofold. 
In the absence of solvent, NQEs destabilize the folded state relative to the unfolded state. 
Contrary to the prevailing view that secondary NQEs mainly affect hydrogen bonding, we find significant contributions from the destabilization of hydrophobic motifs, which have so far been overlooked. 
Secondly, simulating \ce{H} $\to$ \ce{D} substitution of the active hydrogen, we find that intrinsic NQEs generally stabilize the peptide, consistent with the experimentally observed isotope effects on protein stability, showing the potentially important role of intrinsic NQEs. 
Together, these findings show that NQEs can substantially influence peptide stability to the same extent as, or even more than, solvent-moderated NQEs, and can account for the experimentally observed general stabilization of proteins upon deuteration.

\section{Results}

For this study, we carefully selected a total of six representative peptide structures.
The first two among them are the {\textalpha}-helix (\ref{fig:structures}a) and {\textbeta}-sheet hairpin (\ref{fig:structures}b) conformations of
capped deca-polyalanine (Ace-Ala\textsubscript{10}-Nme), referred to as ALA-10. 
Despite its simplicity, this model peptide, with its relatively rigid backbone and lack of side-groups, allows us to systematically estimate the impact of NQEs in backbone hydrogen-bonding interactions on the stability of distinct protein secondary structures.
To examine more realistic peptides with complex side-chain interactions and diverse hydrogen-bonding patterns, we also selected four peptide fragments with short HBs (which imply strong hydrogen bonds~\cite{Zhou_ChemSci_2019_v10_p7734}), extracted from Protein Data Bank (PDB) structures~\cite{pdb_rcsb,pdb_rcsb_new} following the procedure detailed in the SI.
The selected fragments are: 
frag-1, a 6-residue sequence (indices 1249–1254) from PDB 5QI6 that adopts a random coil-like conformation (\ref{fig:structures}c); 
frag-2, a 7-residue sequence (indices 106–112) from PDB 3BWH that presents as a coil with a partial helix (\ref{fig:structures}d); 
frag-3, a 7-residue sequence (indices 76–82) from PDB 6F54 that forms a hairpin structure (\ref{fig:structures}e); 
and frag-4, a 8-residue sequence (indices 57–64) from PDB 6HMD as a coil containing a beta-turn (\ref{fig:structures}f).

\begin{figure}
    \centering
    \includegraphics[width=0.9\textwidth]{figures/structure.pdf}
    \caption{
    Optimized structures of the peptide fragments investigated in this work. 
    (a) ALA-10 in {\textalpha}-helical conformation and (b) {\textbeta}-hairpin conformation. 
    Protein fragment structures from the Protein Data Bank: 
    (c) frag-1 (from 5QI6~\cite{5qib_pdb}), 
    (d) frag-2 (from 3BWH~\cite{3bwh_pdb}), 
    (e) frag-3 (from 6F54~\cite{6f54_pdb}), and 
    (f) frag-4 (from 6HMD~\cite{6hmd_pdb}). 
    All structures are shown in stick representation (carbon in light cyan, oxygen in red, nitrogen in blue, sulfur in yellow, hydrogen in white),
    with secondary structure elements highlighted as ribbon representation. 
    Hydrogen bonds that stabilize the secondary structures are shown as yellow dashed lines.
    }
    \label{fig:structures}
\end{figure}

We first obtained the folded structures of these peptide sequences, saturated them with hydrogen atoms, and capped them with acetyl and N-methyl amide groups at the N- and C-terminals respectively. 
These structures were then optimized at the density functional theory (DFT) level, using the {\textomega}B97M-D3(BJ)~\cite{Mardirossian_JChemPhys_2016_v144_p214110,Grimme_ChemRev_2016_v116_p5105} dispersion-corrected range-separated hybrid meta-GGA functional with the def2-TZVPPD basis set \cite{Weigend_PhysChemChemPhys_2005_v7_p3297}, which is known to have excellent performance for organic molecules~\cite{Najibi_JChemTheoryComput_2018_v14_p5725,Santra_AipConfProc_2019}.
To increase efficiency, the structures were first optimized coarsely using the MACE-OFF23(M) potential~\cite{kovacsMACEOFF23TransferableMachine2023a} -- a pretrained model MLIP for organic systems at the same DFT level -- followed by direct DFT optimization.
Note that in all these peptide fragments, the very short heavy-atom separations increased significantly after optimization, from \qtyrange[range-phrase=--]{2.31}{2.48}{\AA} in the crystal structures to \qtyrange[range-phrase=--]{2.67}{2.78}{\AA}.
This elongation of the very short HBs has been reported in a previous study, and occurs even when the backbone atoms of the amino acid pairs are fixed~\cite{Qi2019}.

\begin{figure}
    \centering
    \includegraphics[width=1.0\textwidth]{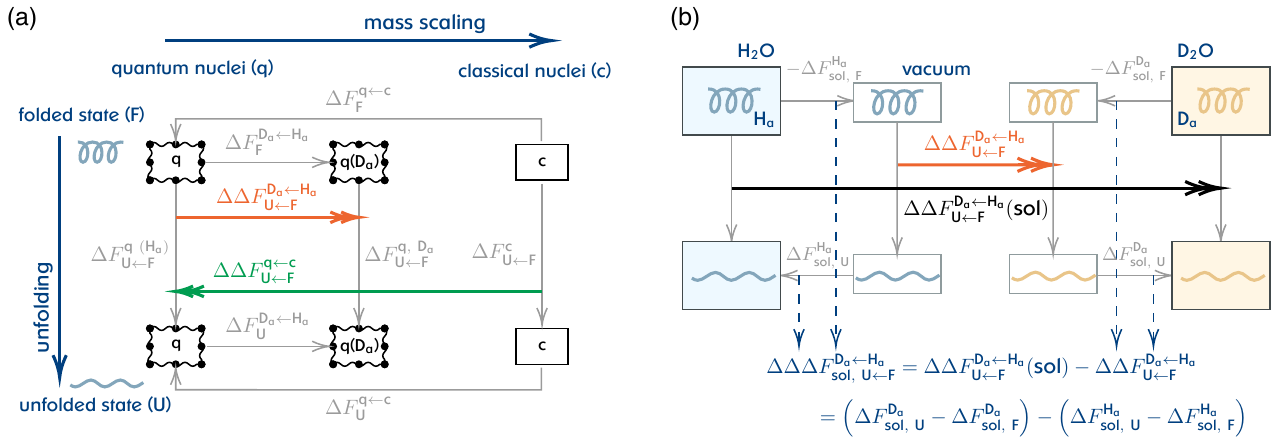}
    \caption{
    Thermodynamic cycles illustrating the free energy relationships in peptide state transitions.
    \textbf{(a)} The gas-phase thermodynamic cycle showing relationships between peptide conformational states with classical and quantum nuclei.
    The vertical dimension represents conformational transitions between folded (F) and unfolded (U) states, while the horizontal dimension represents a continuous transition from quantum nuclei (q) to classical nuclei (c) through mass scaling.
    Boxes circled with ring-polymer beads represent systems with quantum nuclei, and D\textsubscript{a} means the active hydrogens in the system are deuterated. 
    The double arrows represent changes in the folding free energies: the green double arrow represent the change in the folding free energy from classical nuclei to quantum nuclei, and the red double arrow represent the change folding free energy upon deuteration of the active hydrogens.
    \textbf{(b)} Thermodynamic cycle illustrating the partitioning of the experimentally-observed isotope effects on protein stability in solution into intrinsic and solvation contributions.
    The blue and yellow shading of the peptides represent H\textsubscript{a} and D\textsubscript{a} states, respectively.
    The rectangles with transparent, blue and yellow background indicate vacuum, \ce{H2O} and \ce{D2O} environments, respectively.
    The experimentally-observed total free-energy change from \ce{H2O} solvent to \ce{D2O} solvent ($\Delta \Delta F_{\mathrm{U}\leftarrow \mathrm{F}}^{\mathrm{D}_{\mathrm{a}}\leftarrow \mathrm{H}_{\mathrm{a}}}\mathrm{( sol)}$) can be decomposed into two components: 
    (1) the intrinsic peptide isotope effects $\Delta \Delta F_{\mathrm{U}\leftarrow \mathrm{F}}^{\mathrm{D}_{\mathrm{a}}\leftarrow \mathrm{H}_{\mathrm{a}}}$, 
    and (2) the differential solvation free energy $\Delta \Delta \Delta F_{\mathrm{sol} ,\ \mathrm{U}\leftarrow \mathrm{F}}^{\mathrm{D}_{\mathrm{a}}\leftarrow \mathrm{H}_{\mathrm{a}}}$.
    }
    \label{fig:mass-ti-diagram}
\end{figure}

We investigate two key thermodynamic quantities here. 
The first is the change in the folding free energy, which is the free energy difference between the folded peptide and its unfolded state, due to NQEs (illustrated by the green double-arrow in \ref{fig:mass-ti-diagram}a),
\begin{equation}
\Delta \Delta F_{\mathrm{U}\leftarrow \mathrm{F}}^{\mathrm{q}\leftarrow \mathrm{c}} \equiv \Delta F_{\mathrm{U}\leftarrow \mathrm{F}}^{\mathrm{q}} -\Delta F_{\mathrm{U}\leftarrow \mathrm{F}}^{\mathrm{c}} =\Delta F_{\mathrm{U}}^{\mathrm{q\leftarrow c}} -\Delta F_{\mathrm{F}}^{\mathrm{q\leftarrow c}}.
\label{eq:df_c2q}
\end{equation}
The superscripts `q' and `c' stand for `quantum nuclei' and `classical nuclei', respectively; and the subscripts `U' and `F' stand for `unfolded state' and `folded state', respectively.
This quantity directly measures the contribution of NQEs to protein stability, which is of fundamental theoretical importance. 
Experimentally however, NQEs in peptides are commonly measured as isotope effects by comparing peptides solvated in \ce{H2O} and \ce{D2O}, which arises from the difference in folding free energy of the peptide in the two solvents.
This free energy difference can be rigorously decomposed into two components: the intrinsic (i.e. gas-phase) folding free energy change upon deuteration of active hydrogens in the peptide and the solvation free energy differences of the peptide in \ce{H2O} versus \ce{D2O}.
The former component, as depicted by the orange double-arrow in \ref{fig:mass-ti-diagram}, is the second quantity we examine,
\begin{equation}
\Delta \Delta F_{\mathrm{U}\leftarrow \mathrm{F}}^{\mathrm{D_a}\leftarrow \mathrm{H_a}} \equiv \Delta F_{\mathrm{U}\leftarrow \mathrm{F}}^{\mathrm{D_a}} -\Delta F_{\mathrm{U}\leftarrow \mathrm{F}}^{\mathrm{H_a}} =\Delta F_{\mathrm{U}}^{\mathrm{D_a\leftarrow H_a}} -\Delta F_{\mathrm{F}}^{\mathrm{D_a\leftarrow H_a}}.
\label{eq:df_ah2ad}
\end{equation}
The subscript `a' denotes active hydrogens, which are the hydrogens attached to polar heteroatoms (N, O, S) that can be spontaneously deuterated when solvated in \ce{D2O}. 

To compute the above free energy differences rigorously from first principles, thermodynamic integration is carried out in combination with \textit{ab initio} level PIMD simulations, as detailed in the methods section.
PIMD captures NQEs by exploiting the isomorphism between a quantum particle and a classical ring polymer, simulating multiple coupled replicas of the classical system (beads) to represent quantum statistics.
To make these simulations computationally viable, the a new MLIP is transfer learnt from the MACE-OFF23(M) MLIP \cite{kovacsMACEOFF23TransferableMachine2023a} via singlehead finetuning~\cite{Kaur_FaradayDiscuss_2025_v256_p120,gawkowski2025goodbaduglyatomistic} to achieve \textit{ab initio} accuracy, as detailed in the methods section.

\subsection{Impact of nuclear quantum effects on the stability of peptides}
\begin{figure}
    \centering
    \includegraphics[width=0.7\textwidth]{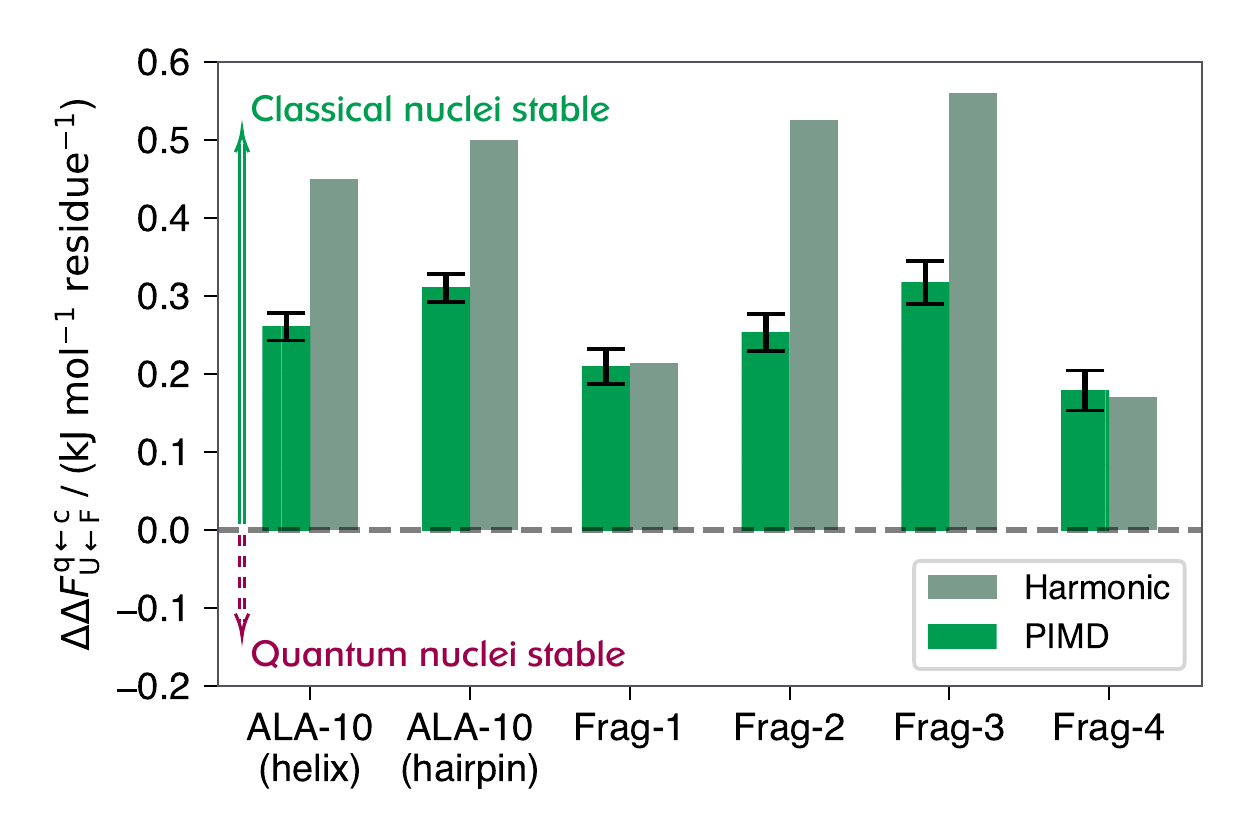}
    \caption{
    Summary of NQEs on the stability ($\Delta \Delta F_{\mathrm{U}\leftarrow \mathrm{F}}^{\mathrm{q}\leftarrow \mathrm{c}}$ per residue) across all studied peptide systems.
    Bright green bars with black error bars represent PIMD simulation results, and dim green bars show harmonic approximation results.
    Positive value indicates NQEs destabilize the peptide's folded state relative to the unfolded state, while negative value indicates stabilization of the folded state due to NQEs.
    }
    \label{fig:nqes}
\end{figure}

We first quantify how much NQEs alter the stability of peptides by estimating the free energy difference in Eq.~\ref{eq:df_c2q} at room temperature.
As shown in \ref{fig:nqes}, NQEs increase the free energy of both the helical and hairpin structures of ALA-10, relative to the unfolded structure, by \qty{0.26 \pm 0.02} and \qty{0.31 \pm 0.02}{kJ\per mol} per residue, respectively.
This means that both helical and hairpin structures of ALA-10 are destabilized by NQEs at room temperature.
Given that the backbone HBs of the ALA-10 peptide are relatively weak, as evidenced by the relatively large heavy atom separations (see SI Section S1), this observation seems to aligns with the CQE picture that NQEs weakens weak HBs~\cite{Li_ProcNatlAcadSciUSA_2011_v108_p6369}.
To quantify the absolute scale of NQEs' impact on the folding free energy, we computed the classical folding free energy using umbrella sampling (see SI Section S3).
Results show that for helical ALA-10, the classical folding free energy is $\sim$\qty{8}{kJ\per mol} per residue, indicating that NQEs reduce it by $\sim$3\%.

Given the destabilizing effect of NQEs observed in the simple ALA-10 peptide with relatively weak hydrogen bonds, a natural question arises as to whether this impact of NQEs changes qualitatively in more complex peptides with side chains and stronger hydrogen bonds.
To this end, we selected four peptide (frag-1 to frag-4) from protein crystal structures, chosen for their diverse secondary motifs and features indicative of stronger hydrogen bonding, namely polar side chains and short heavy-atom separations in crystal structures.
Based on the qualitative CQE picture, which posits that NQEs strengthen strong hydrogen bonds~\cite{Li_ProcNatlAcadSciUSA_2011_v108_p6369}, one would anticipate reduced destabilizing NQEs and possible stabilizing NQEs in these peptides.
However, PIMD simulations reveal surprising results: all four fragments are destabilized by NQEs, with free energy increases of \qtylist[list-units=single]{0.21\pm 0.03; 0.25\pm 0.03; 0.32\pm 0.03; 0.18\pm 0.03}{kJ\per mol} per residue for frag-1 through frag-4, respectively, quantitatively close to the destabilization observed for the ALA10 peptide.
These findings appear to uncover a trend where NQEs generally destabilize peptide structures across diverse structural motifs, irrespective of hydrogen bond strength or the presence of side chains.

To understand the surprising destabilizing trend of NQEs uncovered on the folded peptide configurations, we perform analysis within the harmonic approximation to Eq.~\ref{eq:df_c2q} (see SI Section S4).
Firstly, we benchmark the harmonic approximation against the rigorous PIMD results we have obtained. 
Previous studies have shown that this approximation generally delivers reasonable qualitative performance, even for anharmonic systems, by virtue of error cancellations~\cite{Fang_JPhysChemLett_2016_v7_p2125,Kapil_JChemTheoryComput_2019_v15_p5845}. 
Despite the highly anharmonic nature of peptide systems, \ref{fig:nqes} demonstrates that the harmonic approximation qualitatively predicts the correct destabilizing NQEs across all the systems studied. 
The consistent qualitative agreement indicates that the harmonic approximation captures the essential physics governing NQEs in these peptides.

\begin{figure}
    \centering
    \includegraphics[width=1.0\textwidth]{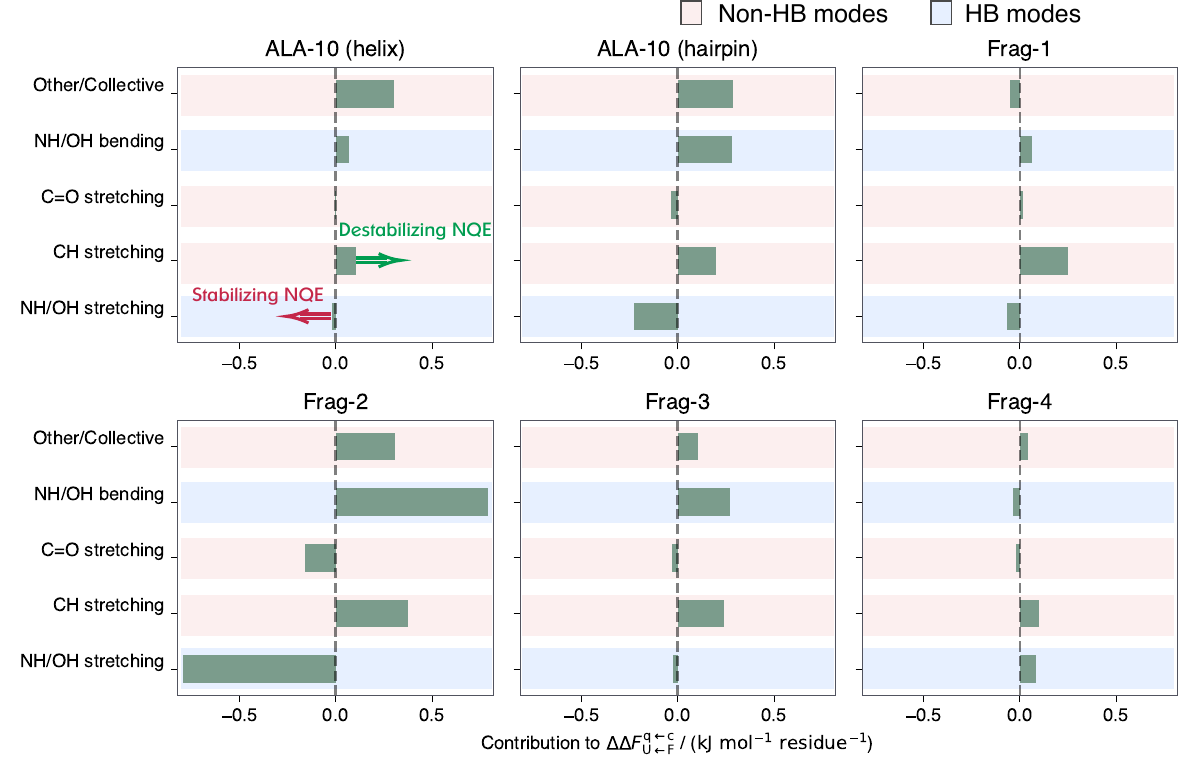}
    \caption{
    Decomposition of the NQEs on the folding free energies in the harmonic limit 
    into contributions from the following modes: NH/OH stretching, CH stretching, C=O stretching, NH/OH bending, and Other/Collective modes. 
    Each panel presents results for a different peptide system studied.
    Negative values indicate NQEs stabilize the folded state relative to the unfolded state, while positive values indicate the opposite.
    Blue background highlights HB vibrational modes, which are emphasized in the conventional CQE model, while pink background highlights non-HB modes.
    }
    \label{fig:mode_decompose}
\end{figure}

A key advantage of the harmonic approximation is that it enables a straightforward decomposition of $\Delta \Delta F_{\mathrm{U}\leftarrow \mathrm{F}}^{\mathrm{q} \leftarrow \mathrm{c}}$ into contributions from individual vibrational modes. 
This can provide crucial insights into the mechanistic origin of the destabilizing NQEs in the peptide systems.
We categorized the vibrational modes in the peptide structures into five groups, 
listed in descending order of their typical frequency range: 
NH/OH stretching, CH stretching, C=O stretching, NH/OH bending, and Other/Collective modes
(see SI Section~S4).
\ref{fig:mode_decompose} shows the decomposition of NQEs on the folding free energy into the above five vibrational mode groups for all six peptides. 
The CQE picture describes NQEs in H-bonded systems as the competition between the NQEs in HB stretching, which strengthen HBs, and the NQEs in HB bending, which weaken HBs \cite{Li_ProcNatlAcadSciUSA_2011_v108_p6369,Ceriotti_ChemRev_2016_v116_p7529}.
The two competing modes correspond to the N–H/O–H stretching and bending mode groups in the decomposition, which are highlighted in blue in \ref{fig:mode_decompose}.  
Consistent with the CQE picture, decomposition shows that the HB bending modes contribute to destabilization while the HB stretching modes contribute to stabilization.
Systems with stronger HBs, e.g. frag-2, show a much larger stabilization NQEs from the HB stretching modes compared to systems with weak HBs, e.g. ALA-10 hairpin conformer.
However, for several peptide fragments, e.g. ALA-10(helix), frag-1, and frag-4, contributions from HB stretching and bending modes are insignificant. 
In frag-4, HB stretching modes surprisingly contribute to destabilization, likely due to the presence of HBs in the unfolded structure.

\ref{fig:mode_decompose} clearly demonstrates that the conventional CQE picture --- 
which predominantly focus on HB modes
--- is inadequate for describing NQEs in complex biomolecules. 
Importantly, beyond the contribution from HB modes, NQEs in the C-H stretching modes, which are often overlooked in the conventional CQE picture, also play a substantial role in destabilization across all the folded peptides studied.
Free hydrogens in C-H bonds experience a more confined environment in the folded peptide, leading to slightly higher vibrational frequencies (see SI). 
Given the large number of these modes, their cumulative effect becomes substantial. 
In additional, NQEs in low-frequency collective modes generally contribute to destabilizing the folded structure. 
Thus, we propose a generalized CQE picture for complex biomolecules encompasses competing effects within HBs, destabilizing contributions from low-frequency collective modes, and, most critically, destabilization driven by C-H stretching. 
Our findings underscore that while H-bonding undoubtedly underpins protein secondary structures, an exclusive focus on NQEs in HBs risks qualitatively misleading predictions for the stability of complex H-bonded systems, where NQEs in C–H stretching often dominate.

\subsection{Impact of isotope  substitution of active protons on the stability of peptides}
\begin{figure}
    \centering
    \includegraphics[width=1.0\textwidth]{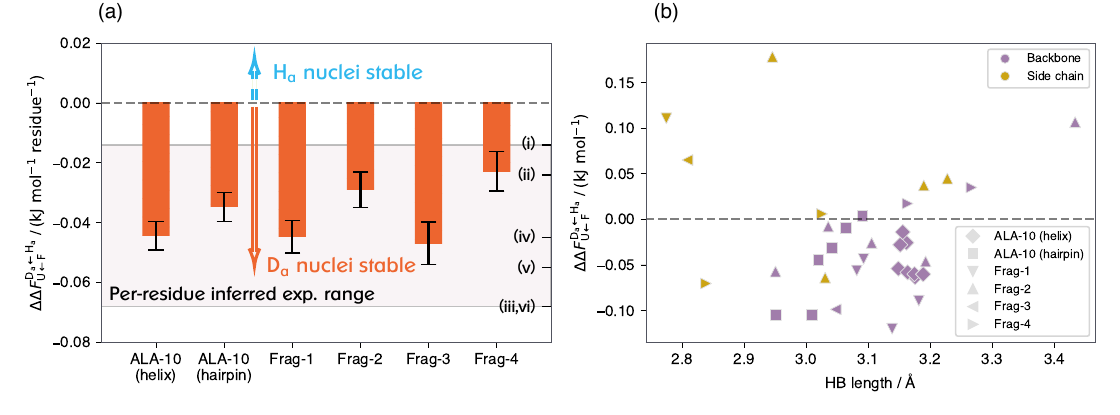}
    \caption{
    H/D isotope effects on peptide conformational stability.
    \textbf{(a)} Folding free energy change upon deuteration of the active hydrogens ($\Delta \Delta F_{\mathrm{U}\leftarrow \mathrm{F}}^{\mathrm{D_a}\leftarrow \mathrm{H_a}}$) calculated from PIMD for the six peptides. 
    The range of experimental total isotope effects is shown in shaded region, with literature values marked for different protein/peptide systems: 
    (i) Equine heart cytochrome c~\cite{krantz2000d}, 
    (ii) Cross-linked GCN4~\cite{krantz2000d}, 
    (iii) Hen egg lysozyme~\cite{Efimova_Biopolymers_2007_v85_p264}, 
    (iv) Bovine serum albumin~\cite{Efimova_Biopolymers_2007_v85_p264}, 
    (v) Rat CD2 domain 1~\cite{Parker_Biochemistry_1997_v36_p5786}, 
    (vi) SH3 domain~\cite{Stadmiller_ProteinSciPublProteinSoc_2018_v27_p1710}.
    \textbf{(b)} Contributions of each individual active H that forms a stable HB in the folded conformer, plotted against trajectory-averaged HB length (heavy-atom separation distance).
    Points are colored to distinguish backbone (purple) from side-chain (yellow) HBs. 
    Marker shapes represent different peptides: diamond for ALA-10 (helix), square for ALA-10 (hairpin), downward triangle for frag-1, upward triangle for frag-2, leftward triangle for frag-3, and rightward triangle for frag-4.
    }
    \label{fig:isotope}
\end{figure}

Having established the overall impact of NQEs in destabilizing the folded peptides, we now examine their experimentally accessible counterpart: 
the change in the stability of a peptide arising from isotope substitution of the solvent ~\cite{Makhatadze_NatStructBiol_1995_v2_p852,Chellgren_JAmChemSoc_2004_v126_p14734,Cho_JAmChemSoc_2009_v131_p15188,Stadmiller_ProteinSciPublProteinSoc_2018_v27_p1710}.
Substituting the \ce{H2O} solvent with \ce{D2O} not only alters the solvation environment of the peptide, but also substitutes exchangeable (active) hydrogen atoms, i.e. the N-H and O-H hydrogens. 
Therefore, the experimentally measurable free-energy change upon \ce{H2O} $\to$ \ce{D2O} solvent substitution ($\Delta \Delta F_{\mathrm{U}\leftarrow \mathrm{F}}^{\mathrm{D}_{\mathrm{a}}\leftarrow \mathrm{H}_{\mathrm{a}}} \ (\mathrm{sol})$) can thus be separated into two contributions (\ref{fig:mass-ti-diagram}b): 
(1) intrinsic isotope effects from substitution of active hydrogens ($\Delta \Delta F_{\mathrm{U}\leftarrow \mathrm{F}}^{\mathrm{D}_{\mathrm{a}}\leftarrow \mathrm{H}_{\mathrm{a}}}$), which modifies the relative stability of the folded versus unfolded states of the peptide, and
(2) solvation free energy differences between the peptide in \ce{H2O} and the peptide with active hydrogen deuterated in \ce{D2O} ($\Delta \Delta \Delta F_{\mathrm{sol} ,\ \mathrm{U}\leftarrow \mathrm{F}}^{\mathrm{D}_{\mathrm{a}}\leftarrow \mathrm{H}_{\mathrm{a}}}$).
In \ref{fig:isotope}a, we report the former contribution by computing Eq.~\ref{eq:df_ah2ad} from PIMD simulations for both forms of ALA-10 and frag-1 through frag-4.
We find that deuterium substitution decreases the free energies by \num{0.044 \pm 0.005} and \qty{0.034 \pm 0.005}{kJ\per mol} per residue for helical and hairpin ALA-10 conformers, respectively, relative to the unfolded state.
Similarly, for frag-1 through frag-4, deuteration of the active hydrogens results in decreases of the relative free energy of the folded state by \num{0.045 \pm 0.006}, \num{0.029 \pm 0.006}, \num{0.047 \pm 0.008} and \qty{0.023 \pm 0.007}{kJ\per mol} per residue, respectively.
The consistent negative values indicate that deuteration stabilizes the folded structures relative to their unfolded states across all systems examined.
These results agree with our preceding analyses, as deuteration increases the nuclear mass, thereby attenuating destabilizing NQEs.
These findings are also consistent with the experimental observations that the majority of the peptides become more stable in \ce{D2O}.

On an absolute scale these numbers are small, and at face value one might consider this contribution negligible. 
However, it is instructive to compare these values with experimentally measured isotope effects on the stability of solvated proteins~\cite{Parker_Biochemistry_1997_v36_p5786,krantz2000d,Efimova_Biopolymers_2007_v85_p264,Stadmiller_ProteinSciPublProteinSoc_2018_v27_p1710}. 
These free energy differences typically range from a fraction to several kcal/mol; thus, when normalized by the number of residues, these values correspond to the shaded region in \ref{fig:isotope}a.
For example, the N-terminal SH3 domain of the \textit{Drosophila} drk protein shows a stabilization of nearly \qty{1}{kcal\per mol} in \ce{D2O}~\cite{Stadmiller_ProteinSciPublProteinSoc_2018_v27_p1710}, corresponding to approximately \qty{0.07}{kJ\per mol} per residue.
Intriguingly, the per residue experimental results on the total free energy change upon solvent iostopic substitution ($\Delta \Delta F_{\mathrm{U}\leftarrow \mathrm{F}}^{\mathrm{D}_{\mathrm{a}}\leftarrow \mathrm{H}_{\mathrm{a}}}\mathrm{( sol)}$) align quantitatively with our calculated intrinsic stabilization free energies associated with active hydrogen deuteration ($\Delta \Delta F_{\mathrm{U}\leftarrow \mathrm{F}}^{\mathrm{D}_{\mathrm{a}}\leftarrow \mathrm{H}_{\mathrm{a}}}$).
Even accounting for incomplete deuterium exchange due to residue burial~\cite{Efimova_Biopolymers_2007_v85_p264}, the resulting intrinsic peptide isotope effects remain significant enough to contribute substantially to the total observed stabilization effect.
This finding challenges the prevailing view that attributes isotope effects predominantly to changes in hydrophobic interactions or bulk solvent properties~\cite{Giubertoni2023}, demonstrating that intrinsic changes in peptide hydrogen bonding represent a significant, rather than negligible, contribution.

To elucidate the microscopic origins of this stabilization,
we decomposed the total isotope contribution by examining individual sites based on their local environments and hydrogen bonding characteristics (\ref{fig:isotope}b and SI Section S5).
While no clear correlations emerged with standard descriptors such as HB length or backbone dihedral angles (see SI),
an interesting trend emerges when separating HBs into backbone and side-chain categories.
Deuteration of backbone amide hydrogens predominantly leads to stabilization of the folded structure.
In contrast, deuteration of active side-chain hydrogens exhibits more diverse behaviors, ranging from modest stabilization to sizable destabilization.
This heterogeneity implies that while the global trend favors stabilization after deuteration,
consistent with the vast majority of the experimental observations, destabilization may occur in peptides dominated by side-chain hydrogen bonding.

\section{Conclusion}
To summarize, we answer the fundamental question of how NQEs impact the stability of peptides using PIMD simulations combined with MLIP trained and fine-tuned on \textit{ab initio} data at the range-separated hybrid functional DFT level.
Our simulations reveal that NQEs destabilize secondary peptide structures across different motifs by approximately \qty{0.2}{kJ\per mol} per amino acid. 
Further analysis show that the conventional competing quantum effects picture -- centered on HB stretching versus bending -- is inadequate for describing complex bio-molecular systems: NQEs in the non-hydrogen-bonded C–H stretching modes play a key role in destabilizing folded structures.
Isotope substitution calculations uncover that deuteration of the active hydrogens generally stabilizes peptides by $\sim$ 0.04 kJ/mol per residue. 
Quantitatively, this isotope effects intrinsic to the peptide is comparable to the experimentally measured free energy difference between peptides solvated in \ce{H2O} and \ce{D2O}.
This indicates that in contrast to conventional views that focus on solvent isotope effects, intrinsic isotopes effects associated with the deuteration of the active peptide hydrogens also plays a crucial role in the experimentally observed stabilization of folded peptides in \ce{D2O}.
Further analysis show that deuteration of backbone active hydrogens predominately result in stabilization of the folded peptide.
Our study delivers the long-desired microscopic insights on the role of NQEs in peptides, paving the way for rational prediction and design of isotopic substitution strategies in practical applications.

\section{Methods and Computational Details}
\subsection{Fine-tuning MACE-OFF potential}
System-specific MLIPs were developed by fine-tuning the pre-trained MACE-OFF23(M) foundation model for each peptide system.
Training datasets of 300--800 configurations were generated through targeted molecular dynamics sampling at multiple temperatures, with structures selected to span diverse conformational space (see SI Section S2 for details).
All reference electronic structure calculations for energies and forces were performed at the {\textomega}B97M-D3(BJ)/def2-TZVPPD level of theory using the ORCA quantum chemistry package~\cite{ORCA,ORCA6}, maintaining consistency with the protocol of the MACE-OFF23 model.
The final fine-tuned models achieved root-mean-square errors of \qty{< 0.035}{kJ\per mol} per atom for energies and \qty{< 2.7}{kJ\per mol\per \AA} for forces across all systems. 
Performance metrics for all systems are provided in SI Tables S2–S6.

\subsection{Thermodynamic integration from classical to quantum}
We implemented a comprehensive thermodynamic integration over mass scaling (mass-TI) to quantify the contribution of NQEs to peptide stability, as illustrated in \ref{fig:mass-ti-diagram}.
The mass-TI approach has been commonly employed for the simulation of H/D isotope effects~\cite{Perez_JChemTheoryComput_2011_v7_p2358,Ceriotti_JChemPhys_2013_v138_p14112}.
This approach has also been extended to simulate classical to quantum free energy changes, becoming a well-established method that has been widely applied ~\cite{Rossi2015,Fang_JPhysChemLett_2016_v7_p2125,Rossi_PhysRevLett_2016_v117_p115702}.

The free energy change from classical nuclei to quantum nuclei ($\Delta F_{\alpha }^{\mathrm{q}\leftarrow \mathrm{c}}$) of a peptide at a given conformational state $\alpha$ (folded (F) or unfolded (U)) is calculated via thermodynamic integration over mass.
A series of PIMD simulations are performed with the mass of the system scaled by a factor of $\mu$, and $\Delta F_{\alpha }^{\mathrm{q}\leftarrow \mathrm{c}}$ is given by, 
\begin{equation}
\Delta F_{\alpha }^{\mathrm{q}\leftarrow \mathrm{c}} =2\int _{0}^{1}\frac{\langle K( \lambda ) \rangle ^{\mathrm{q}} -\langle K\rangle ^{\mathrm{c}}}{\lambda }\mathrm{d} \lambda ,
\label{Fcq}
\end{equation}
where $\lambda = \mu ^{-1/2}$, $\langle K\rangle ^{\mathrm{q}}$ and $\displaystyle \langle K\rangle ^{\mathrm{c}}=\tfrac{3}{2}N_\text{a}k_\text{B}T$ are the average kinetic energies of the system in the quantum and classical ensembles, respectively; $N_\text{a}$ is the number of atoms.
Similarly, the free energy difference between a system with deuterium (D) and protium (H) can be computed via PIMD simulations scaling the mass of H,
\begin{equation}
\Delta F_{\alpha }^{\mathrm{D}\leftarrow \mathrm{H}} =2\int _{1}^{2^{-1/2}}\frac{\langle K_{\mathrm{H}}( \lambda ) \rangle ^{\mathrm{q}} -\langle K_{\mathrm{H}} \rangle ^{\mathrm{c}}}{\lambda }\mathrm{d} \lambda ,
\label{FHD}
\end{equation}
where $\langle K_{\mathrm{H}} \rangle ^{\mathrm{q}}$ and $\langle K_{\mathrm{H}} \rangle ^{\mathrm{c}}=\tfrac{3}{2}N_\text{H}k_\text{B}T$ are the quantum and classical kinetic energies of all the isotope-substituted H atoms (the active H atoms), respectively. 
We used 11 and 4 evenly spaced integration points to numerically evaluate the integrals in Eq.~\ref{Fcq} and Eq.~\ref{FHD}, respectively.

All PIMD simulations are performed in the canonical ensemble ($NVT$) at \qty{300}{K}.
The PIGLET thermostat~\cite{Ceriotti_PhysRevLett_2012_v109_p100604} is employed to reduce the number of beads required to converge the quantum kinetic energy, achieving excellent convergence with only 8 beads.
The equations of motion were integrated with a time step of \qty{0.5}{fs}. 
To achieve better sampling, for each peptide conformation, we run four independent trajectories from different initial structures.
Each simulation underwent \qty{5}{ps} of equilibration followed by \qty{100}{ps} of production, with the latter extended to \qty{300}{ps} for isotope effects calculations to ensure sufficient statistical precision. 
All simulations were performed using the i-PI 3.0 code~\cite{Litman_JChemPhys_2024_v161_p062504} interfaced with our fine-tuned MACE models.
Convergence of the PIMD simulations has been carefully examined and presented in the SI.

\begin{acknowledgments}
The authors thank Prof. Tong Zhu for the helpful discussions and advices on the project.
This work is supported by the National Natural Science Foundation of China (grant number 22373021).                       
\end{acknowledgments}

\bibliography{ref}

\begin{thebibliography}{10}
\expandafter\ifx\csname url\endcsname\relax
  \def\url#1{\texttt{#1}}\fi
\expandafter\ifx\csname urlprefix\endcsname\relax\def\urlprefix{URL }\fi
\providecommand{\bibinfo}[2]{#2}
\providecommand{\eprint}[2][]{\url{#2}}

\bibitem{Anfinsen_SciNewYorkNY_1973_v181_p223}
\bibinfo{author}{Anfinsen, C.~B.}
\newblock \bibinfo{title}{{Principles that govern the folding of protein
  chains}}.
\newblock \emph{\bibinfo{journal}{Science}} \textbf{\bibinfo{volume}{181}},
  \bibinfo{pages}{223--30} (\bibinfo{year}{1973}).

\bibitem{Dobson_Nature_2003_v426_p884}
\bibinfo{author}{Dobson, C.~M.}
\newblock \bibinfo{title}{{Protein folding and misfolding}}.
\newblock \emph{\bibinfo{journal}{Nature}} \textbf{\bibinfo{volume}{426}},
  \bibinfo{pages}{884--90} (\bibinfo{year}{2003}).

\bibitem{Jumper_Nature_2021_v596_p583}
\bibinfo{author}{Jumper, J.} \emph{et~al.}
\newblock \bibinfo{title}{{Highly accurate protein structure prediction with
  AlphaFold}}.
\newblock \emph{\bibinfo{journal}{Nature}} \textbf{\bibinfo{volume}{596}},
  \bibinfo{pages}{583--589} (\bibinfo{year}{2021}).

\bibitem{Dill_Biochemistry_1990_v29_p7133}
\bibinfo{author}{Dill, K.~A.}
\newblock \bibinfo{title}{{Dominant forces in protein folding}}.
\newblock \emph{\bibinfo{journal}{Biochemistry}} \textbf{\bibinfo{volume}{29}},
  \bibinfo{pages}{7133--55} (\bibinfo{year}{1990}).

\bibitem{Newberry_AcsChemBiol_2019_v14_p1677}
\bibinfo{author}{Newberry, R.~W.} \& \bibinfo{author}{Raines, R.~T.}
\newblock \bibinfo{title}{{Secondary Forces in Protein Folding}}.
\newblock \emph{\bibinfo{journal}{Acs Chem. Biol.}}
  \textbf{\bibinfo{volume}{14}}, \bibinfo{pages}{1677--1686}
  (\bibinfo{year}{2019}).

\bibitem{PAULING_ProcNatlAcadSciUSA_1951_v37_p205}
\bibinfo{author}{PAULING, L.}, \bibinfo{author}{COREY, R.~B.} \&
  \bibinfo{author}{BRANSON, H.~R.}
\newblock \bibinfo{title}{{The structure of proteins; two hydrogen-bonded
  helical configurations of the polypeptide chain}}.
\newblock \emph{\bibinfo{journal}{Proc. Natl. Acad. Sci. U. S. A.}}
  \textbf{\bibinfo{volume}{37}}, \bibinfo{pages}{205--11}
  (\bibinfo{year}{1951}).

\bibitem{Pauling_ProcNatlAcadSciUSA_1951_v37_p729}
\bibinfo{author}{Pauling, L.} \& \bibinfo{author}{Corey, R.~B.}
\newblock \bibinfo{title}{{Configurations of Polypeptide Chains With Favored
  Orientations Around Single Bonds: Two New Pleated Sheets}}.
\newblock \emph{\bibinfo{journal}{Proc. Natl. Acad. Sci. U. S. A.}}
  \textbf{\bibinfo{volume}{37}}, \bibinfo{pages}{729--40}
  (\bibinfo{year}{1951}).

\bibitem{Hudgins_JAmChemSoc_1999_v121_p3494}
\bibinfo{author}{Hudgins, R.~R.} \& \bibinfo{author}{Jarrold, M.~F.}
\newblock \bibinfo{title}{{Helix Formation in Unsolvated Alanine-Based
  Peptides:{\,} Helical Monomers and Helical Dimers}}.
\newblock \emph{\bibinfo{journal}{J. Am. Chem. Soc.}}
  \textbf{\bibinfo{volume}{121}}, \bibinfo{pages}{3494--3501}
  (\bibinfo{year}{1999}).

\bibitem{Rose_AnnuRevBiophysBiomolStruct_1993_v22_p381}
\bibinfo{author}{Rose, G.~D.} \& \bibinfo{author}{Wolfenden, R.}
\newblock \bibinfo{title}{{Hydrogen bonding, hydrophobicity, packing, and
  protein folding}}.
\newblock \emph{\bibinfo{journal}{Annu. Rev. Biophys. Biomol. Struct.}}
  \textbf{\bibinfo{volume}{22}}, \bibinfo{pages}{381--415}
  (\bibinfo{year}{1993}).

\bibitem{Pace_ProteinSciPublProteinSoc_2014_v23_p652}
\bibinfo{author}{Pace, C.~N.} \emph{et~al.}
\newblock \bibinfo{title}{{Contribution of hydrogen bonds to protein
  stability}}.
\newblock \emph{\bibinfo{journal}{Protein Sci.: Publ. Protein Soc.}}
  \textbf{\bibinfo{volume}{23}}, \bibinfo{pages}{652--61}
  (\bibinfo{year}{2014}).

\bibitem{Li_ProcNatlAcadSciUSA_2011_v108_p6369}
\bibinfo{author}{Li, X.-Z.}, \bibinfo{author}{Walker, B.} \&
  \bibinfo{author}{Michaelides, A.}
\newblock \bibinfo{title}{{Quantum nature of the hydrogen bond}}.
\newblock \emph{\bibinfo{journal}{Proc. Natl. Acad. Sci. U.S.A.}}
  \textbf{\bibinfo{volume}{108}}, \bibinfo{pages}{6369--6373}
  (\bibinfo{year}{2011}).

\bibitem{DiMartino_NatRevDrugDiscov_2023_v22_p562}
\bibinfo{author}{{Di Martino}, R. M.~C.}, \bibinfo{author}{Maxwell, B.~D.} \&
  \bibinfo{author}{Pirali, T.}
\newblock \bibinfo{title}{{Deuterium in drug discovery: progress, opportunities
  and challenges}}.
\newblock \emph{\bibinfo{journal}{Nat. Rev., Drug Discov.}}
  \textbf{\bibinfo{volume}{22}}, \bibinfo{pages}{562--584}
  (\bibinfo{year}{2023}).

\bibitem{Cherrah_BiomedEnvironMassSpectrom_1987_v14_p653}
\bibinfo{author}{Cherrah, Y.} \emph{et~al.}
\newblock \bibinfo{title}{{Study of deuterium isotope effects on protein
  binding by gas chromatography/mass spectrometry. Caffeine and deuterated
  isotopomers}}.
\newblock \emph{\bibinfo{journal}{Biomed. Environ. Mass Spectrom.}}
  \textbf{\bibinfo{volume}{14}}, \bibinfo{pages}{653--7}
  (\bibinfo{year}{1987}).

\bibitem{Parente_FoodChemToxicolIntJPublBrIndBiolResAssoc_2022_v160_p112774}
\bibinfo{author}{Parente, R.~M.}, \bibinfo{author}{Tarantino, P.~M.},
  \bibinfo{author}{Sippy, B.~C.} \& \bibinfo{author}{Burdock, G.~A.}
\newblock \bibinfo{title}{{Pharmacokinetic, pharmacological, and genotoxic
  evaluation of deuterated caffeine}}.
\newblock \emph{\bibinfo{journal}{Food Chem. Toxicol.: Int. J. Publ. Br. Ind.
  Biol. Res. Assoc.}} \textbf{\bibinfo{volume}{160}}, \bibinfo{pages}{112774}
  (\bibinfo{year}{2022}).

\bibitem{el1984lipophilicity}
\bibinfo{author}{El~Tayar, N.}, \bibinfo{author}{van~de Waterbeemd, H.},
  \bibinfo{author}{Gryllaki, M.}, \bibinfo{author}{Testa, B.} \&
  \bibinfo{author}{Trager, W.~F.}
\newblock \bibinfo{title}{The lipophilicity of deuterium atoms. a comparison of
  shake-flask and hplc methods}.
\newblock \emph{\bibinfo{journal}{International journal of pharmaceutics}}
  \textbf{\bibinfo{volume}{19}}, \bibinfo{pages}{271--281}
  (\bibinfo{year}{1984}).

\bibitem{Bechalany_HelvChimActa_1989_v72_p472}
\bibinfo{author}{Bechalany, A.} \emph{et~al.}
\newblock \bibinfo{title}{{Isotope Effects on the Lipophilicity of Deuterated
  Caffeines}}.
\newblock \emph{\bibinfo{journal}{Helv. Chim. Acta}}
  \textbf{\bibinfo{volume}{72}}, \bibinfo{pages}{472--476}
  (\bibinfo{year}{1989}).

\bibitem{Schnauss_AdvMater_2021_v33_p2170230}
\bibinfo{author}{Schnau{\ss}, J.} \emph{et~al.}
\newblock \bibinfo{title}{{Cells in Slow Motion: Cells in Slow Motion: Apparent
  Undercooling Increases Glassy Behavior at Physiological Temperatures (Adv.
  Mater. 29/2021)}}.
\newblock \emph{\bibinfo{journal}{Adv. Mater.}} \textbf{\bibinfo{volume}{33}},
  \bibinfo{pages}{2170230} (\bibinfo{year}{2021}).

\bibitem{Jandova_Cancers_2021_v13_p605}
\bibinfo{author}{Jandova, J.}, \bibinfo{author}{Hua, A.~B.},
  \bibinfo{author}{Fimbres, J.} \& \bibinfo{author}{Wondrak, G.~T.}
\newblock \bibinfo{title}{{Deuterium Oxide (D2O) Induces Early Stress Response
  Gene Expression and Impairs Growth and Metastasis of Experimental Malignant
  Melanoma}}.
\newblock \emph{\bibinfo{journal}{Cancers}} \textbf{\bibinfo{volume}{13}},
  \bibinfo{pages}{605} (\bibinfo{year}{2021}).

\bibitem{Giubertoni2023}
\bibinfo{author}{Giubertoni, G.}, \bibinfo{author}{Bonn, M.} \&
  \bibinfo{author}{Woutersen, S.}
\newblock \bibinfo{title}{D2o as an imperfect replacement for h2o: Problem or
  opportunity for protein research?}
\newblock \emph{\bibinfo{journal}{The Journal of Physical Chemistry B}}
  \textbf{\bibinfo{volume}{127}}, \bibinfo{pages}{8086–8094}
  (\bibinfo{year}{2023}).
\newblock \urlprefix\url{http://dx.doi.org/10.1021/acs.jpcb.3c04385}.

\bibitem{Parker_Biochemistry_1997_v36_p5786}
\bibinfo{author}{Parker, M.~J.} \& \bibinfo{author}{Clarke, A.~R.}
\newblock \bibinfo{title}{{Amide backbone and water-related H/D isotope effects
  on the dynamics of a protein folding reaction}}.
\newblock \emph{\bibinfo{journal}{Biochemistry}} \textbf{\bibinfo{volume}{36}},
  \bibinfo{pages}{5786--94} (\bibinfo{year}{1997}).

\bibitem{Chellgren_JAmChemSoc_2004_v126_p14734}
\bibinfo{author}{Chellgren, B.~W.} \& \bibinfo{author}{Creamer, T.~P.}
\newblock \bibinfo{title}{{Effects of H2O and D2O on polyproline II helical
  structure}}.
\newblock \emph{\bibinfo{journal}{J. Am. Chem. Soc.}}
  \textbf{\bibinfo{volume}{126}}, \bibinfo{pages}{14734--5}
  (\bibinfo{year}{2004}).

\bibitem{Cho_JAmChemSoc_2009_v131_p15188}
\bibinfo{author}{Cho, Y.} \emph{et~al.}
\newblock \bibinfo{title}{{Hydrogen bonding of beta-turn structure is
  stabilized in D(2)O}}.
\newblock \emph{\bibinfo{journal}{J. Am. Chem. Soc.}}
  \textbf{\bibinfo{volume}{131}}, \bibinfo{pages}{15188--93}
  (\bibinfo{year}{2009}).

\bibitem{Makhatadze_NatStructBiol_1995_v2_p852}
\bibinfo{author}{Makhatadze, G.~I.}, \bibinfo{author}{Clore, G.~M.} \&
  \bibinfo{author}{Gronenborn, A.~M.}
\newblock \bibinfo{title}{{Solvent isotope effect and protein stability}}.
\newblock \emph{\bibinfo{journal}{Nat. Struct. Biol.}}
  \textbf{\bibinfo{volume}{2}}, \bibinfo{pages}{852--5} (\bibinfo{year}{1995}).

\bibitem{Ceriotti_ChemRev_2016_v116_p7529}
\bibinfo{author}{Ceriotti, M.} \emph{et~al.}
\newblock \bibinfo{title}{{Nuclear Quantum Effects in Water and Aqueous
  Systems: Experiment, Theory, and Current Challenges}}.
\newblock \emph{\bibinfo{journal}{Chem. Rev.}} \textbf{\bibinfo{volume}{116}},
  \bibinfo{pages}{7529--50} (\bibinfo{year}{2016}).

\bibitem{Haidar_JAmSocMassSpectrom_2023_v34_p1447}
\bibinfo{author}{Haidar, Y.} \& \bibinfo{author}{Konermann, L.}
\newblock \bibinfo{title}{{Effects of Hydrogen/Deuterium Exchange on Protein
  Stability in Solution and in the Gas Phase}}.
\newblock \emph{\bibinfo{journal}{J. Am. Soc. Mass Spectrom.}}
  \textbf{\bibinfo{volume}{34}}, \bibinfo{pages}{1447--1458}
  (\bibinfo{year}{2023}).

\bibitem{Benoit_Nature_1998_v392_p258}
\bibinfo{author}{Benoit, M.}, \bibinfo{author}{Marx, D.} \&
  \bibinfo{author}{Parrinello, M.}
\newblock \bibinfo{title}{{Tunnelling and zero-point motion in high-pressure
  ice}}.
\newblock \emph{\bibinfo{journal}{Nature}} \textbf{\bibinfo{volume}{392}},
  \bibinfo{pages}{258--261} (\bibinfo{year}{1998}).

\bibitem{Chen_PhysRevLett_2003_v91_p215503}
\bibinfo{author}{Chen, B.}, \bibinfo{author}{Ivanov, I.},
  \bibinfo{author}{Klein, M.~L.} \& \bibinfo{author}{Parrinello, M.}
\newblock \bibinfo{title}{{Hydrogen bonding in water}}.
\newblock \emph{\bibinfo{journal}{Phys. Rev. Lett.}}
  \textbf{\bibinfo{volume}{91}}, \bibinfo{pages}{215503}
  (\bibinfo{year}{2003}).

\bibitem{Morrone_PhysRevLett_2008_v101_p17801}
\bibinfo{author}{Morrone, J.~A.} \& \bibinfo{author}{Car, R.}
\newblock \bibinfo{title}{{Nuclear quantum effects in water}}.
\newblock \emph{\bibinfo{journal}{Phys. Rev. Lett.}}
  \textbf{\bibinfo{volume}{101}}, \bibinfo{pages}{17801}
  (\bibinfo{year}{2008}).

\bibitem{Paesani_JPhysChemB_2009_v113_p5702}
\bibinfo{author}{Paesani, F.} \& \bibinfo{author}{Voth, G.~A.}
\newblock \bibinfo{title}{{The properties of water: insights from quantum
  simulations}}.
\newblock \emph{\bibinfo{journal}{J. Phys. Chem., B}}
  \textbf{\bibinfo{volume}{113}}, \bibinfo{pages}{5702--19}
  (\bibinfo{year}{2009}).

\bibitem{Wang_ProcNatlAcadSciUSA_2014_v111_p18454}
\bibinfo{author}{Wang, L.}, \bibinfo{author}{Fried, S.~D.},
  \bibinfo{author}{Boxer, S.~G.} \& \bibinfo{author}{Markland, T.~E.}
\newblock \bibinfo{title}{{Quantum delocalization of protons in the
  hydrogen-bond network of an enzyme active site}}.
\newblock \emph{\bibinfo{journal}{Proc. Natl. Acad. Sci. U. S. A.}}
  \textbf{\bibinfo{volume}{111}}, \bibinfo{pages}{18454--9}
  (\bibinfo{year}{2014}).

\bibitem{Rossi2015}
\bibinfo{author}{Rossi, M.}, \bibinfo{author}{Fang, W.} \&
  \bibinfo{author}{Michaelides, A.}
\newblock \bibinfo{title}{Stability of complex biomolecular structures: van der
  waals, hydrogen bond cooperativity, and nuclear quantum effects}.
\newblock \emph{\bibinfo{journal}{The Journal of Physical Chemistry Letters}}
  \textbf{\bibinfo{volume}{6}}, \bibinfo{pages}{4233–4238}
  (\bibinfo{year}{2015}).
\newblock \urlprefix\url{http://dx.doi.org/10.1021/acs.jpclett.5b01899}.

\bibitem{Fang_JPhysChemLett_2016_v7_p2125}
\bibinfo{author}{Fang, W.} \emph{et~al.}
\newblock \bibinfo{title}{{Inverse Temperature Dependence of Nuclear Quantum
  Effects in DNA Base Pairs}}.
\newblock \emph{\bibinfo{journal}{J. Phys. Chem. Lett.}}
  \textbf{\bibinfo{volume}{7}}, \bibinfo{pages}{2125--31}
  (\bibinfo{year}{2016}).

\bibitem{Rossi_PhysRevLett_2016_v117_p115702}
\bibinfo{author}{Rossi, M.}, \bibinfo{author}{Gasparotto, P.} \&
  \bibinfo{author}{Ceriotti, M.}
\newblock \bibinfo{title}{{Anharmonic and Quantum Fluctuations in Molecular
  Crystals: A First- Principles Study of the Stability of Paracetamol}}.
\newblock \emph{\bibinfo{journal}{Phys. Rev. Lett.}}
  \textbf{\bibinfo{volume}{117}}, \bibinfo{pages}{115702}
  (\bibinfo{year}{2016}).

\bibitem{Kapil_ProcNatlAcadSciUSA_2022_v119_pe2111769119}
\bibinfo{author}{Kapil, V.} \& \bibinfo{author}{Engel, E.~A.}
\newblock \bibinfo{title}{{A complete description of thermodynamic stabilities
  of molecular crystals}}.
\newblock \emph{\bibinfo{journal}{Proc. Natl. Acad. Sci. U. S. A.}}
  \textbf{\bibinfo{volume}{119}}, \bibinfo{pages}{e2111769119}
  (\bibinfo{year}{2022}).

\bibitem{DellaPia2025}
\bibinfo{author}{Della~Pia, F.} \emph{et~al.}
\newblock \bibinfo{title}{Accurate and efficient machine learning interatomic
  potentials for finite temperature modelling of molecular crystals}.
\newblock \emph{\bibinfo{journal}{Chemical Science}}
  \textbf{\bibinfo{volume}{16}}, \bibinfo{pages}{11419–11433}
  (\bibinfo{year}{2025}).
\newblock \urlprefix\url{http://dx.doi.org/10.1039/D5SC01325A}.

\bibitem{Pereyaslavets_ProcNatlAcadSciUSA_2018_v115_p8878}
\bibinfo{author}{Pereyaslavets, L.} \emph{et~al.}
\newblock \bibinfo{title}{{On the importance of accounting for nuclear quantum
  effects in ab initio calibrated force fields in biological simulations}}.
\newblock \emph{\bibinfo{journal}{Proc. Natl. Acad. Sci. U. S. A.}}
  \textbf{\bibinfo{volume}{115}}, \bibinfo{pages}{8878--8882}
  (\bibinfo{year}{2018}).

\bibitem{Ugur2025}
\bibinfo{author}{Ugur, B.~E.} \& \bibinfo{author}{Webb, M.~A.}
\newblock \bibinfo{title}{Nuclear quantum effects in molecular liquids across
  chemical space}.
\newblock \emph{\bibinfo{journal}{Nature Communications}}
  \textbf{\bibinfo{volume}{16}} (\bibinfo{year}{2025}).
\newblock \urlprefix\url{http://dx.doi.org/10.1038/s41467-025-60850-x}.

\bibitem{Errea_Nature_2016_v532_p81}
\bibinfo{author}{Errea, I.} \emph{et~al.}
\newblock \bibinfo{title}{{Quantum hydrogen-bond symmetrization in the
  superconducting hydrogen sulfide system}}.
\newblock \emph{\bibinfo{journal}{Nature}} \textbf{\bibinfo{volume}{532}},
  \bibinfo{pages}{81--4} (\bibinfo{year}{2016}).

\bibitem{Li_PhysRevLett_2010_v104_p66102}
\bibinfo{author}{Li, X.-Z.}, \bibinfo{author}{Probert, M. I.~J.},
  \bibinfo{author}{Alavi, A.} \& \bibinfo{author}{Michaelides, A.}
\newblock \bibinfo{title}{{Quantum nature of the proton in water-hydroxyl
  overlayers on metal surfaces}}.
\newblock \emph{\bibinfo{journal}{Phys. Rev. Lett.}}
  \textbf{\bibinfo{volume}{104}}, \bibinfo{pages}{66102}
  (\bibinfo{year}{2010}).

\bibitem{Goncharov_SciNewYorkNY_1996_v273_p218}
\bibinfo{author}{Goncharov, A.}, \bibinfo{author}{Struzhkin, V.},
  \bibinfo{author}{Somayazulu, M.}, \bibinfo{author}{Hemley, R.} \&
  \bibinfo{author}{Mao, H.}
\newblock \bibinfo{title}{{Compression of Ice to 210 Gigapascals: Infrared
  Evidence for a Symmetric Hydrogen-Bonded Phase}}.
\newblock \emph{\bibinfo{journal}{Science}} \textbf{\bibinfo{volume}{273}},
  \bibinfo{pages}{218--20} (\bibinfo{year}{1996}).

\bibitem{Ubbelohde_ActaCryst_1955_v8_p71}
\bibinfo{author}{Ubbelohde, A.~R.} \& \bibinfo{author}{Gallagher, K.~J.}
\newblock \bibinfo{title}{{Acid-base effects in hydrogen bonds in crystals}}.
\newblock \emph{\bibinfo{journal}{Acta Cryst}} \textbf{\bibinfo{volume}{8}},
  \bibinfo{pages}{71--83} (\bibinfo{year}{1955}).

\bibitem{Shi_NatCommun_2018_v9_p481}
\bibinfo{author}{Shi, C.}, \bibinfo{author}{Zhang, X.}, \bibinfo{author}{Yu,
  C.-H.}, \bibinfo{author}{Yao, Y.-F.} \& \bibinfo{author}{Zhang, W.}
\newblock \bibinfo{title}{{Geometric isotope effect of deuteration in a
  hydrogen-bonded host- guest crystal}}.
\newblock \emph{\bibinfo{journal}{Nat. Commun.}} \textbf{\bibinfo{volume}{9}},
  \bibinfo{pages}{481} (\bibinfo{year}{2018}).

\bibitem{Habershon_JChemPhys_2009_v131_p24501}
\bibinfo{author}{Habershon, S.}, \bibinfo{author}{Markland, T.~E.} \&
  \bibinfo{author}{Manolopoulos, D.~E.}
\newblock \bibinfo{title}{{Competing quantum effects in the dynamics of a
  flexible water model}}.
\newblock \emph{\bibinfo{journal}{J. Chem. Phys.}}
  \textbf{\bibinfo{volume}{131}}, \bibinfo{pages}{24501}
  (\bibinfo{year}{2009}).

\bibitem{Markland_ProcNatlAcadSciUSA_2012_v109_p7988}
\bibinfo{author}{Markland, T.~E.} \& \bibinfo{author}{Berne, B.~J.}
\newblock \bibinfo{title}{{Unraveling quantum mechanical effects in water using
  isotopic fractionation}}.
\newblock \emph{\bibinfo{journal}{Proc. Natl. Acad. Sci. U. S. A.}}
  \textbf{\bibinfo{volume}{109}}, \bibinfo{pages}{7988--91}
  (\bibinfo{year}{2012}).

\bibitem{Romanelli_JPhysChemLett_2013_v4_p3251}
\bibinfo{author}{Romanelli, G.} \emph{et~al.}
\newblock \bibinfo{title}{{Direct Measurement of Competing Quantum Effects on
  the Kinetic Energy of Heavy Water upon Melting}}.
\newblock \emph{\bibinfo{journal}{J. Phys. Chem. Lett.}}
  \textbf{\bibinfo{volume}{4}}, \bibinfo{pages}{3251--3256}
  (\bibinfo{year}{2013}).

\bibitem{Ceriotti_ProcNatlAcadSciUSA_2013_v110_p15591}
\bibinfo{author}{Ceriotti, M.}, \bibinfo{author}{Cuny, J.},
  \bibinfo{author}{Parrinello, M.} \& \bibinfo{author}{Manolopoulos, D.~E.}
\newblock \bibinfo{title}{{Nuclear quantum effects and hydrogen bond
  fluctuations in water}}.
\newblock \emph{\bibinfo{journal}{Proc. Natl. Acad. Sci. U. S. A.}}
  \textbf{\bibinfo{volume}{110}}, \bibinfo{pages}{15591--6}
  (\bibinfo{year}{2013}).

\bibitem{Cheng_JPhysChemLett_2016_v7_p2210}
\bibinfo{author}{Cheng, B.}, \bibinfo{author}{Behler, J.} \&
  \bibinfo{author}{Ceriotti, M.}
\newblock \bibinfo{title}{{Nuclear Quantum Effects in Water at the Triple
  Point: Using Theory as a Link Between Experiments}}.
\newblock \emph{\bibinfo{journal}{J. Phys. Chem. Lett.}}
  \textbf{\bibinfo{volume}{7}}, \bibinfo{pages}{2210--5}
  (\bibinfo{year}{2016}).

\bibitem{Zhou_ChemSci_2019_v10_p7734}
\bibinfo{author}{Zhou, S.} \& \bibinfo{author}{Wang, L.}
\newblock \bibinfo{title}{{Unraveling the structural and chemical features of
  biological short hydrogen bonds}}.
\newblock \emph{\bibinfo{journal}{Chem. Sci.}} \textbf{\bibinfo{volume}{10}},
  \bibinfo{pages}{7734--7745} (\bibinfo{year}{2019}).

\bibitem{pdb_rcsb}
\bibinfo{author}{Berman, H.~M.} \emph{et~al.}
\newblock \bibinfo{title}{{The Protein Data Bank}}.
\newblock \emph{\bibinfo{journal}{Nucleic Acids Res.}}
  \textbf{\bibinfo{volume}{28}}, \bibinfo{pages}{235--42}
  (\bibinfo{year}{2000}).
\newblock \urlprefix\url{https://www.rcsb.org}.

\bibitem{pdb_rcsb_new}
\bibinfo{author}{Burley, S.~K.} \emph{et~al.}
\newblock \bibinfo{title}{{Updated resources for exploring
  experimentally-determined PDB structures and Computed Structure Models at the
  RCSB Protein Data Bank}}.
\newblock \emph{\bibinfo{journal}{Nucleic Acids Res.}}
  \textbf{\bibinfo{volume}{53}}, \bibinfo{pages}{D564--D574}
  (\bibinfo{year}{2025}).

\bibitem{5qib_pdb}
\bibinfo{author}{Schuller, M.} \emph{et~al.}
\newblock \bibinfo{title}{{PanDDA analysis group deposition -- Crystal
  Structure of human PARP14 Macrodomain 3 in complex with FMOPL000597a}}.
\newblock \emph{\bibinfo{journal}{Worldw. Protein Data Bank}}
  (\bibinfo{year}{2019}).

\bibitem{3bwh_pdb}
\bibinfo{author}{Chen, L.}
\newblock \bibinfo{title}{{Atomic resolution structure of cucurmosin, a novel
  type 1 RIP from the sarcocarp of Cucurbita moschata}}.
\newblock \emph{\bibinfo{journal}{Worldw. Protein Data Bank}}
  (\bibinfo{year}{2008}).

\bibitem{6f54_pdb}
\bibinfo{author}{Dunstan, M.} \& \bibinfo{author}{Currin, A.}
\newblock \bibinfo{title}{{CRYSTAL STRUCTURE OF KETOSTEROID ISOMERASE TRIPLE
  VARIANT V88I/L99VD103S}}.
\newblock \emph{\bibinfo{journal}{Worldw. Protein Data Bank}}
  (\bibinfo{year}{2018}).

\bibitem{6hmd_pdb}
\bibinfo{author}{Niefind, K.}, \bibinfo{author}{Lindenblatt, D.},
  \bibinfo{author}{Jose, J.} \& \bibinfo{author}{{Le Borgne}, M.}
\newblock \bibinfo{title}{{STRUCTURE OF PROTEIN KINASE CK2 CATALYTIC SUBUNIT
  (ISOFORM CK2ALPHA'; CSNK2A2 gene product) IN COMPLEX WITH THE
  INDENOINDOLE-TYPE INHIBITOR AR18}}.
\newblock \emph{\bibinfo{journal}{Worldw. Protein Data Bank}}
  (\bibinfo{year}{2019}).

\bibitem{Mardirossian_JChemPhys_2016_v144_p214110}
\bibinfo{author}{Mardirossian, N.} \& \bibinfo{author}{Head-Gordon, M.}
\newblock \bibinfo{title}{{{\ensuremath{\omega}}B97M-V: A combinatorially
  optimized, range- separated hybrid, meta-GGA density functional with VV10
  nonlocal correlation}}.
\newblock \emph{\bibinfo{journal}{J. Chem. Phys.}}
  \textbf{\bibinfo{volume}{144}}, \bibinfo{pages}{214110}
  (\bibinfo{year}{2016}).

\bibitem{Grimme_ChemRev_2016_v116_p5105}
\bibinfo{author}{Grimme, S.}, \bibinfo{author}{Hansen, A.},
  \bibinfo{author}{Brandenburg, J.~G.} \& \bibinfo{author}{Bannwarth, C.}
\newblock \bibinfo{title}{{Dispersion-Corrected Mean-Field Electronic Structure
  Methods}}.
\newblock \emph{\bibinfo{journal}{Chem. Rev.}} \textbf{\bibinfo{volume}{116}},
  \bibinfo{pages}{5105--54} (\bibinfo{year}{2016}).

\bibitem{Weigend_PhysChemChemPhys_2005_v7_p3297}
\bibinfo{author}{Weigend, F.} \& \bibinfo{author}{Ahlrichs, R.}
\newblock \bibinfo{title}{{Balanced basis sets of split valence, triple zeta
  valence and quadruple zeta valence quality for H to Rn: Design and assessment
  of accuracy}}.
\newblock \emph{\bibinfo{journal}{Phys. Chem. Chem. Phys.}}
  \textbf{\bibinfo{volume}{7}}, \bibinfo{pages}{3297} (\bibinfo{year}{2005}).

\bibitem{Najibi_JChemTheoryComput_2018_v14_p5725}
\bibinfo{author}{Najibi, A.} \& \bibinfo{author}{Goerigk, L.}
\newblock \bibinfo{title}{{The Nonlocal Kernel in van der Waals Density
  Functionals as an Additive Correction: An Extensive Analysis with Special
  Emphasis on the B97M-V and {\ensuremath{\omega}}B97M-V Approaches}}.
\newblock \emph{\bibinfo{journal}{J. Chem. Theory Comput.}}
  \textbf{\bibinfo{volume}{14}}, \bibinfo{pages}{5725--5738}
  (\bibinfo{year}{2018}).

\bibitem{Santra_AipConfProc_2019}
\bibinfo{author}{Santra, G.} \& \bibinfo{author}{Martin, J. M.~L.}
\newblock \bibinfo{title}{{Some observations on the performance of the most
  recent exchange- correlation functionals for the large and chemically diverse
  GMTKN55 benchmark}} (\bibinfo{year}{2019}).

\bibitem{kovacsMACEOFF23TransferableMachine2023a}
\bibinfo{author}{Kov{\'a}cs, D.~P.} \emph{et~al.}
\newblock \bibinfo{title}{{{MACE-OFF23}}: {{Transferable Machine Learning Force
  Fields}} for {{Organic Molecules}}} (\bibinfo{year}{2023}).
\newblock \urlprefix\url{https://arxiv.org/abs/2312.15211}.
\newblock \eprint{2312.15211}.

\bibitem{Qi2019}
\bibinfo{author}{Qi, H.~W.} \& \bibinfo{author}{Kulik, H.~J.}
\newblock \bibinfo{title}{Evaluating unexpectedly short non-covalent distances
  in x-ray crystal structures of proteins with electronic structure analysis}.
\newblock \emph{\bibinfo{journal}{Journal of Chemical Information and
  Modeling}} \textbf{\bibinfo{volume}{59}}, \bibinfo{pages}{2199–2211}
  (\bibinfo{year}{2019}).
\newblock \urlprefix\url{http://dx.doi.org/10.1021/acs.jcim.9b00144}.

\bibitem{Kaur_FaradayDiscuss_2025_v256_p120}
\bibinfo{author}{Kaur, H.} \emph{et~al.}
\newblock \bibinfo{title}{{Data-efficient fine-tuning of foundational models
  for first-principles quality sublimation enthalpies}}.
\newblock \emph{\bibinfo{journal}{Faraday Discuss.}}
  \textbf{\bibinfo{volume}{256}}, \bibinfo{pages}{120--138}
  (\bibinfo{year}{2025}).

\bibitem{gawkowski2025goodbaduglyatomistic}
\bibinfo{author}{Gawkowski, M.~J.}, \bibinfo{author}{Li, M.},
  \bibinfo{author}{Shi, B.~X.} \& \bibinfo{author}{Kapil, V.}
\newblock \bibinfo{title}{The good, the bad, and the ugly of atomistic learning
  for "clusters-to-bulk" generalization} (\bibinfo{year}{2025}).
\newblock \urlprefix\url{https://arxiv.org/abs/2509.16601}.
\newblock \eprint{2509.16601}.

\bibitem{Kapil_JChemTheoryComput_2019_v15_p5845}
\bibinfo{author}{Kapil, V.}, \bibinfo{author}{Engel, E.},
  \bibinfo{author}{Rossi, M.} \& \bibinfo{author}{Ceriotti, M.}
\newblock \bibinfo{title}{{Assessment of Approximate Methods for Anharmonic
  Free Energies}}.
\newblock \emph{\bibinfo{journal}{J. Chem. Theory Comput.}}
  \textbf{\bibinfo{volume}{15}}, \bibinfo{pages}{5845--5857}
  (\bibinfo{year}{2019}).

\bibitem{Stadmiller_ProteinSciPublProteinSoc_2018_v27_p1710}
\bibinfo{author}{Stadmiller, S.~S.} \& \bibinfo{author}{Pielak, G.~J.}
\newblock \bibinfo{title}{{Enthalpic stabilization of an SH3 domain by D2 O}}.
\newblock \emph{\bibinfo{journal}{Protein Sci.: Publ. Protein Soc.}}
  \textbf{\bibinfo{volume}{27}}, \bibinfo{pages}{1710--1716}
  (\bibinfo{year}{2018}).

\bibitem{krantz2000d}
\bibinfo{author}{Krantz, B.~A.}, \bibinfo{author}{Moran, L.~B.},
  \bibinfo{author}{Kentsis, A.} \& \bibinfo{author}{Sosnick, T.~R.}
\newblock \bibinfo{title}{D/h amide kinetic isotope effects reveal when
  hydrogen bonds form during protein folding}.
\newblock \emph{\bibinfo{journal}{nature structural biology}}
  \textbf{\bibinfo{volume}{7}}, \bibinfo{pages}{62--71} (\bibinfo{year}{2000}).

\bibitem{Efimova_Biopolymers_2007_v85_p264}
\bibinfo{author}{Efimova, Y.~M.}, \bibinfo{author}{Haemers, S.},
  \bibinfo{author}{Wierczinski, B.}, \bibinfo{author}{Norde, W.} \&
  \bibinfo{author}{{van Well}, A.~A.}
\newblock \bibinfo{title}{{Stability of globular proteins in H2O and D2O}}.
\newblock \emph{\bibinfo{journal}{Biopolymers}} \textbf{\bibinfo{volume}{85}},
  \bibinfo{pages}{264--73} (\bibinfo{year}{2007}).

\bibitem{ORCA}
\bibinfo{author}{Neese, F.}
\newblock \bibinfo{title}{The orca program system}.
\newblock \emph{\bibinfo{journal}{Wiley Interdiscip. Rev. Comput. Mol. Sci.}}
  \textbf{\bibinfo{volume}{2}}, \bibinfo{pages}{73--78} (\bibinfo{year}{2012}).

\bibitem{ORCA6}
\bibinfo{author}{Neese, F.}
\newblock \bibinfo{title}{Software update: The orca program system—version
  6.0}.
\newblock \emph{\bibinfo{journal}{Wiley Interdiscip. Rev. Comput. Mol. Sci.}}
  \textbf{\bibinfo{volume}{15}}, \bibinfo{pages}{e70019}
  (\bibinfo{year}{2025}).

\bibitem{Perez_JChemTheoryComput_2011_v7_p2358}
\bibinfo{author}{P{\'e}rez, A.} \& \bibinfo{author}{{von Lilienfeld}, O.~A.}
\newblock \bibinfo{title}{{Path Integral Computation of Quantum Free Energy
  Differences Due to Alchemical Transformations Involving Mass and Potential}}.
\newblock \emph{\bibinfo{journal}{J. Chem. Theory Comput.}}
  \textbf{\bibinfo{volume}{7}}, \bibinfo{pages}{2358--69}
  (\bibinfo{year}{2011}).

\bibitem{Ceriotti_JChemPhys_2013_v138_p14112}
\bibinfo{author}{Ceriotti, M.} \& \bibinfo{author}{Markland, T.~E.}
\newblock \bibinfo{title}{{Efficient methods and practical guidelines for
  simulating isotope effects}}.
\newblock \emph{\bibinfo{journal}{J. Chem. Phys.}}
  \textbf{\bibinfo{volume}{138}}, \bibinfo{pages}{14112}
  (\bibinfo{year}{2013}).

\bibitem{Ceriotti_PhysRevLett_2012_v109_p100604}
\bibinfo{author}{Ceriotti, M.} \& \bibinfo{author}{Manolopoulos, D.~E.}
\newblock \bibinfo{title}{{Efficient first-principles calculation of the
  quantum kinetic energy and momentum distribution of nuclei}}.
\newblock \emph{\bibinfo{journal}{Phys. Rev. Lett.}}
  \textbf{\bibinfo{volume}{109}}, \bibinfo{pages}{100604}
  (\bibinfo{year}{2012}).

\bibitem{Litman_JChemPhys_2024_v161_p062504}
\bibinfo{author}{Litman, Y.} \emph{et~al.}
\newblock \bibinfo{title}{{i-PI 3.0: A flexible and efficient framework for
  advanced atomistic simulations}}.
\newblock \emph{\bibinfo{journal}{J. Chem. Phys.}}
  \textbf{\bibinfo{volume}{161}}, \bibinfo{pages}{062504}
  (\bibinfo{year}{2024}).

\end{thebibliography}

\end{document}